\documentclass[aps]{revtex4}
\usepackage{amsmath}
\usepackage{graphicx}
\usepackage[usenames]{color}

\newcommand\beq{ \begin{eqnarray} }
\newcommand\eeq{ \end{eqnarray} }

\begin{document}

\title{BEC-BCS Crossover in the $\epsilon $ Expansion}
\author{Jiunn-Wei Chen}
\author{Eiji Nakano}
\affiliation{
Department of Physics and Center for Theoretical Sciences,
National Taiwan University, Taipei 10617, Taiwan}

\begin{abstract}
The $\epsilon $ expansion (expansion around four spacial dimensions)
developed by Nishida and Son for a cold fermi gas with infinite scattering
length is extended to finite scattering length to study the BEC-BCS
crossover. A resummation of higher order logarithms and a suitable extension
of fermion coupling in d-dimensions\ are developed in order to apply the
theory in the BCS regime. The ratio between the chemical potential and the
Fermi energy, $\mu /\varepsilon _{F}$, is computed to next-to-leading order
in the $\epsilon $ expansion as a function of $\eta =1/(ak_{F})$, where $a$
is the scattering length and $k_{F}$ is the Fermi momentum in a
non-interacting system. Near the unitarity limit $\left\vert \eta
\right\vert \rightarrow 0$, we found $\mu /\varepsilon _{F}=0.475-0.707\eta
-0.5\eta ^{2}$. Near the BEC limit $\eta \rightarrow \infty $, $\mu
/\varepsilon _{F}=0.062/\eta -\eta ^{2}$, while near the BCS limit $\eta
\rightarrow -\infty $, $\mu /\varepsilon _{F}=1+0.707/\eta $. Overall good
agreement with Quantum Monte Carlo results is found.
\end{abstract}

\maketitle



\section{Introduction}

BEC-BCS crossover is a field attracting lots of attention recently \cite%
{AJL1,NSL1,Rand1,OH1,CROSS1}. The simplest systems to study the crossover
are dilute Fermion systems with attractive interactions, for which the
effective range for two-body scattering is much less than the inter-particle
spacing. At zero temperature, such systems are characterized by a
dimensionless number $\eta =1/(ak_{F})$, where $a$ is the two-body
scattering length, and $k_{F}$ is the corresponding Fermi momentum in
non-interacting systems. For $\eta $ large and negative (weak attraction)
one finds the BCS solution with pairing and superfluidity. With $\eta $
large and positive, corresponding to strong attraction with a two-body bound
state well below threshold, the bound pairs will Bose-Einstein condense
(BEC). Experimentally, $\eta $ can be changed between the BEC and BCS limits
at will using the technique of Feshbach resonance \cite%
{OHara,Jin,Grimm,Ketterle,Thomas,Salomon,Thomas05} 
Theoretically, physical quantities are expected to be a smooth function of $%
\eta $, since there is no phase transition between the BEC and BCS limits.
In each of these limits the behavior is nonperturbative, however, the
effective interaction is weak; the system can be successfully described in a
mean field approximation. On the other hand, dilute fermion systems near the
unitarity limit (the $\eta =0$ limit) require treatments beyond the mean
field approximation such as numerical simulations \cite%
{QMC1,QMC2,JWK1,Wingate,CR1,BDM1,Lee}.

Recently, Nishida and Son have proposed an analytical approach called
\textquotedblleft $\epsilon $ expansion\textquotedblright\ to deal with
problems near the unitarity limit \cite{NS1,NS2}. In this approach one
computes physical quantities in $d$ spacial dimensions, then treats $%
\epsilon =4-d$ as a small expansion parameter, and then sets $\epsilon =1$
at the end to obtain the three dimensional physics. It is similar to the
dimensional expansion method used by Wilson to study critical exponents of
second order phase transitions \cite{WK1}. Nishida and Son's idea was
inspired by the observation of Nussinov and Nussinov \cite{NN1} that the
ground state of a two-component fermion system in the unitarity limit is a
free Bose gas for $d\geq 4$ (also see Refs. \cite{Steele,SKC1} for
simplification of certain diagrams with $d\rightarrow \infty $). It is
because the two Fermion bound state (with zero binding energy) wave function
behaves as $\Phi (r)\propto 1/r^{d-2}$, where\ $r$ is the separation between
the two Fermions. The probability integral $\int d^{d}r\left\vert \Phi
(r)\right\vert ^{2}$ has a singularity at $r\rightarrow 0$ when $d\geq 4$.
Thus the bound state has zero size and will not interact with each other.
This feature was explicitly implemented in the theory set up near the
unitarity limit in Refs. \cite{NS1,NS2}. As we will see, away from the
unitarity limit, one also expects the localization of the bound state wave
function all the way to the BEC limit but not to the BCS limit.

Now the $\epsilon $ expansion has been applied to few-body scattering \cite%
{RUP1}, polarized fermions (i.e., with uneven chemical potential between two
fermion species) in the unitarity limit \cite{RSK1}, and near the unitarity
limit \cite{NS2} to identify the \textquotedblleft splitting
point\textquotedblright\ in the phase diagram where three different phases
meet\ \cite{SS1}. The critical temperature in the unitarity limit has also
been computed \cite{Nishida1}. Recently, the first next-to-next-to-leading
order (NNLO) calculation in the $\epsilon $ expansion was carried out for $%
\mu /\varepsilon _{F}$ in the unitarity limit, where $\mu $ is the Fermion
chemical potential and $\varepsilon _{F}$ is the Fermi energy in
non-interacting systems \cite{ADS1}. Contrary to the nice convergence seen
previously at next-to-leading order (NLO), the NNLO of $\mu /\varepsilon
_{F} $ is as large as the leading order (LO) with an opposite sign. Even so,
it is still premature to claim that the large NNLO correction cannot be
tamed through reorganizing the series. Thus, we consider it is still worth
while to explore the $\epsilon $ expansion at lower orders.

In this work, we apply the $\epsilon $ expansion to NLO to study the BEC-BCS
transition with an arbitrary $\eta $. In the BEC side, an interesting
relation between $\mu $ and the two fermion binding energy $B$ is
implemented: 
\begin{equation}
2\mu +B=\mathcal{O}(\epsilon ).  \label{eq1}
\end{equation}%
This comes from the chemical equilibrium between two fermions (with chemical
potential $2\mu $) and one boson (with chemical potential $-B$) in a 4-d
system where the bosons do not interact with each other. We also perform the
resummation of the $\left( \epsilon \log \eta \right) ^{n}/n!$ terms arising
from $\eta ^{\epsilon }-1$. These logarithms are formally higher order in $%
\epsilon $ but numerically important near the unitarity limit\ ($\epsilon =1$%
, $\left\vert \eta \right\vert \ll 1$), and near the BEC and BCS limits ($%
\epsilon =1$, $\left\vert \eta \right\vert \gg 1$). For example, this
resummation gives the linear term in $\mu /\varepsilon _{F}$ when $%
\left\vert \eta \right\vert \ll 1$: 
\begin{equation}
\frac{\mu }{\varepsilon _{F}}=0.475-0.707\eta -0.5\eta ^{2}.
\end{equation}%
Without this resummation, physical quantities are always even in $\eta $
which would prevent us from reaching the BCS side.


\section{$\protect\epsilon $ expansion}

We are interested in a two-component Fermi gas with zero range attractive
interactions characterized by the scattering length $a$. In Nambu-Gor'kov
formalism, the two-component fermions are denoted as $\Psi =(\psi _{\uparrow
},\psi _{\downarrow }^{\dagger })^{T}$. After the Hubbard-Stratonovich
transformation, the Lagrangian of the system is 
\begin{equation}
L=\Psi ^{\dagger }\left( i\partial _{t}+\frac{\nabla ^{2}}{2m}\sigma
_{3}+\mu \sigma _{3}\right) \Psi +\Psi ^{\dagger }\sigma _{+}\Psi \phi +\Psi
^{\dagger }\sigma _{-}\Psi \phi ^{\ast }-\frac{\phi ^{\ast }\phi }{c_{0}},
\label{EQ1}
\end{equation}%
where $m$ is the fermion mass, $\sigma _{i}$ is the Pauli matrix acting on
the spin space, and the coupling $c_{0}$ depends on the two-body scattering
length $a$. The ground state is a superfluid state with a condensate $%
\langle \phi \rangle =\phi _{0}$, whose phase is chosen to be zero. $\phi $
is then expanded around $\phi _{0}$, 
\begin{equation}
\phi =\phi _{0}+g\varphi .  \label{phi_0}
\end{equation}%
The Lagrangian will be further decomposed into perturbative and
non-perturbative parts. We will first look at the BEC side, than the BCS
side.

\subsection{Power Counting in the BEC Side}

The Lagrangian can be decomposed as $L=L_{0}+L_{1}+L_{2}$, 
\begin{eqnarray}
L_{0} &=&\Psi ^{\dagger }\left[ i\partial _{t}+\left( \frac{\nabla ^{2}}{2m}-%
\frac{B}{2}\right) \sigma _{3}+\phi _{0}\sigma _{+}+\phi _{0}\sigma _{-}%
\right] \Psi +\varphi ^{\ast }\left( i\partial _{t}+\frac{\nabla ^{2}}{4m}%
\right) \varphi -\frac{\phi _{0}^{2}}{c_{0}},  \notag \\
L_{1} &=&\frac{\mu _{B}}{2}\Psi ^{\dagger }\sigma _{3}\Psi +g\Psi ^{\dagger
}\sigma _{+}\Psi \varphi +g\Psi ^{\dagger }\sigma _{-}\Psi \varphi ^{\ast
}+\mu _{B}\varphi ^{\ast }\varphi ,  \notag \\
L_{2} &=&-\varphi ^{\ast }\left( i\partial _{t}+\frac{\nabla ^{2}}{4m}+\frac{%
g^{2}}{c_{0}}\right) \varphi -\mu _{B}\varphi ^{\ast }\varphi -g\frac{\phi
_{0}}{c_{0}}(\varphi ^{\ast }+\varphi ),
\end{eqnarray}%
with $B=1/\left( ma^{2}\right) $. $\mu _{B}=2\mu +B\sim \mathcal{O}(\epsilon
)$ from Eq.~(\ref{eq1}) which will be checked explicitly later. Near the BEC
limit ($\eta \gg 1$), $B$ does not vanish in 4-d, thus $B\sim \mathcal{O}%
(\epsilon ^{0})$ and $\mu =(\mu _{B}-B)/2\sim \mathcal{O}(\epsilon ^{0})$.
Similarly, $\phi _{0}$ does not vanish in 4-d in the BEC side. So $\phi
_{0}\sim \mathcal{O}(\epsilon ^{0})$. $c_{0}$ is determined by demanding the
two fermion scattering amplitude has a $1/(p-i/a)$ pole in momentum $p$ in
arbitrary dimensions \cite{RUP1}: 
\begin{equation}
\frac{1}{c_{0}}=\frac{m}{(4\pi )^{2-\epsilon /2}a^{2-\epsilon }}\Gamma
\left( -1+\frac{\epsilon }{2}\right) .  \label{C0}
\end{equation}%
In $3$-d ($\epsilon =1$), $1/c_{0}=-m/(4\pi a)$ where $a$ could be either
positive (in the BEC side) or negative (in the BCS side). However in $4$-d ($%
\epsilon =0$), the scattering amplitude has a double pole $1/(p^{2}+1/a^{2})$%
. Thus, the sign of $a$ does not play a role in 4-d and the difference
between the BEC and BCS sides becomes ambiguous. Besides, if $a<0$, $1/c_{0}$
picks up a phase when $0<\epsilon <1$. To fix this phase, instead of using
Eq. (\ref{C0}), we will use 
\begin{equation}
\frac{1}{c_{0}}=sign[a]\frac{m}{(4\pi )^{2-\epsilon /2}\left\vert
a\right\vert ^{2-\epsilon }}\Gamma \left( -1+\frac{\epsilon }{2}\right) ,
\label{C0x}
\end{equation}%
and write the $\left\vert a\right\vert $ dependence in $1/c_{0}$ as 
\begin{eqnarray}
\left\vert a\right\vert ^{-2+\epsilon } &=&a^{-2}\left( 1+\epsilon X\right) ,
\notag \\
X &=&\left( \left\vert a\right\vert ^{\epsilon }-1\right) /\epsilon
=\sum\limits_{n=1}^{\infty }\frac{\left( \epsilon \log \left\vert
a\right\vert \right) ^{n}}{\epsilon n!},  \label{resum}
\end{eqnarray}%
such that all the higher order logarithms are resummed and lumped into NLO.
Then $\left\vert a\right\vert ^{-2+\epsilon }\rightarrow 1/\left\vert
a\right\vert $ at NLO after $\epsilon \rightarrow 1$ is taken at the end.
Using this prescription, $1/c_{0}\sim \mathcal{O}(\epsilon ^{-1})$.

$L_{0}$ is the non-perturbative part of the lagrangian which defines the
fermion and boson propagators, 
\begin{eqnarray}
S_{F}(p,B) &=&\frac{-1}{\left\{ p_{0}(1+i0^{+})\right\} ^{2}-E_{p}^{2}}\left[
\begin{array}{cc}
p_{0}+\varepsilon _{p}+B/2 & -\phi _{0} \\ 
-\phi _{0} & p_{0}-\varepsilon _{p}-B/2%
\end{array}%
\right] ,  \label{FP1} \\
D(p) &=&\frac{-1}{p_{0}(1+i0^{+})-\varepsilon _{p}/2},  \label{BP1}
\end{eqnarray}%
where $\varepsilon _{p}=\mathbf{p}^{2}/(2m)$, $E_{p}^{2}=\left( \varepsilon
_{p}+B/2\right) ^{2}+\phi _{0}^{2}$ and $B/2$ plays the role of effective
mass for the fermions. $L_{1}$ gives the perturbative interaction with $%
g\sim \mathcal{O}(\epsilon ^{1/2})$ and $\mu _{B}\sim \mathcal{O}(\epsilon
^{1})$, while $L_{2}$ gives the counterterms for boson self energy and
tadpole diagrams. The boson-fermion coupling $g$ is tuned to 
\begin{equation}
g=\frac{\sqrt{8\pi ^{2}\epsilon }}{m}\left( \frac{m\phi _{0}}{2\pi }\right)
^{\epsilon /4}\sim \mathcal{O}(\epsilon ^{1/2}),
\end{equation}%
such that the boson self energy diagram without(with) an insertion of $\mu
_{B}$ will be canceled by the first(second) term in $L_{2}$ in 4-d. The last
term in $L_{2}$ cancels the leading boson tadpole diagram when $\phi _{0}$
is chosen to minimize its effective potential.

The quantity we will compute in this paper is $\mu /\varepsilon _{F}$.
Following the procedure of Ref. \cite{NS1}, we first compute the effective
potential $V_{\mathrm{eff}}(\phi _{0})$ and find the solution $\overline{%
\phi }_{0}$ that minimizes $V_{\mathrm{eff}}$. From $V_{\mathrm{eff}}$, the
fermion number density is 
\begin{equation}
n_{f}=-\left. \frac{\partial V_{\mathrm{eff}}}{\partial \mu }\right\vert
_{\phi _{0}=\overline{\phi }_{0}}.  \label{a}
\end{equation}%
Then $\varepsilon _{F}$ is 
\begin{equation}
\varepsilon _{F}=\frac{2\pi }{m}\left[ \frac{1}{2}\Gamma \left( 3-\frac{%
\epsilon }{2}\right) n_{f}\right] ^{\frac{2}{4-\epsilon }}.  \label{b}
\end{equation}

The leading diagrams for the effective potential $V_{\mathrm{eff}}(\phi _{0})
$ are listed in Fig.~1.

The tree level diagram in Fig.~1(a) gives a contribution 
\begin{equation}
V_{c_{0}}(\phi _{0})=\frac{\phi _{0}^{2}}{c_{0}}=\frac{B}{2^{2-\frac{%
\epsilon }{2}}}\Gamma \left( -1+\frac{\epsilon }{2}\right) \left[ \epsilon
Q\left( \phi _{0}\right) +1\right] \left( \frac{\phi _{0}m}{2\pi }\right)
^{2-\epsilon /2}\sim \mathcal{O}\left( \epsilon ^{-1}\right) ,
\end{equation}%
where 
\begin{equation}
Q(\phi _{0})=\frac{1}{\epsilon }\left[ sign[a]\left( \frac{\phi _{0}}{B}%
\right) ^{\epsilon /2}-1\right] \sim \mathcal{O}(\epsilon ^{0}).
\end{equation}

The one-loop diagram in Fig.~1(b) yields%
\begin{eqnarray}
V_{\mathrm{1L}}(\phi _{0}) &=&-\int \frac{\mathrm{d}p^{5-\epsilon }}{i(2\pi
)^{5-\epsilon }}\ln \det \left[ S_{F}^{-1}(p,B)\right] \,  \notag \\
&=&\phi _{0}\left( \frac{\phi _{0}m}{2\pi }\right) ^{\frac{4-\epsilon }{2}}%
\frac{\pi \,{\left( \frac{B}{\phi _{0}}\right) }^{1-\epsilon
/2}\,_{2}F_{1}\left( \frac{-2+\epsilon }{4},\frac{\epsilon }{4};2;\frac{%
-4\phi _{0}^{2}}{B^{2}}\right) }{2^{2-\epsilon /2}\,\sin \left( \frac{\pi
\epsilon }{2}\right) \Gamma \left( 2-\epsilon /2\right) } \\
&\sim &\mathcal{O}\left( \epsilon ^{-1}\right) ,
\end{eqnarray}%
where $_{2}F_{1}$ is a hypergeometric function. The one-loop diagram in
Fig.~1(c) with one insertion of $\mu _{B}$ yields%
\begin{equation}
V_{\mathrm{1L\mu _{B}}}(\phi _{0})=-\mu _{B}\frac{\partial V_{\mathrm{1L}%
}(\phi _{0})}{\partial B}\sim \mathcal{O}(1).
\end{equation}%
Also, the two-loop diagram in Fig.~1(d) yields 
\begin{eqnarray}
V_{\mathrm{2L}}(\phi _{0}) &=&g^{2}\int \frac{\mathrm{d}p^{5-\epsilon }\,%
\mathrm{d}q^{5-\epsilon }}{i(2\pi )^{5-\epsilon }i(2\pi )^{5-\epsilon }}%
\mathrm{Tr}\left[ S_{F}(p,B)\sigma _{+}S_{F}(q,B)\sigma _{-}\right] D(p-q)\,
\notag \\
&=&-\tilde{C}\left( B/\phi _{0}\right) \,\epsilon \,\phi _{0}\left( \frac{%
\phi _{0}m}{2\pi }\right) ^{\frac{4-\epsilon }{2}}  \label{Cz} \\
&\sim &\mathcal{O}(\epsilon )\ ,  \notag
\end{eqnarray}%
where $\tilde{C}\left( z\right) $ is a decreasing function of $z$ which
converges numerically even at 4-d. The expression for $\tilde{C}\left(
z\right) $ and its numerical plot are shown in the Appendix. So now we have  
\begin{equation}
V_{\mathrm{eff}}=V_{c_{0}}+V_{\mathrm{1L}}+V_{\mathrm{1L\mu _{B}}}+V_{%
\mathrm{2L}}.
\end{equation}

\begin{figure}[tbp]
\begin{center}
\begin{tabular}{cc}
\resizebox{30mm}{!}{\includegraphics{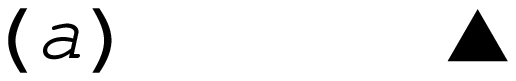}} & \quad %
\resizebox{30mm}{!}{\includegraphics{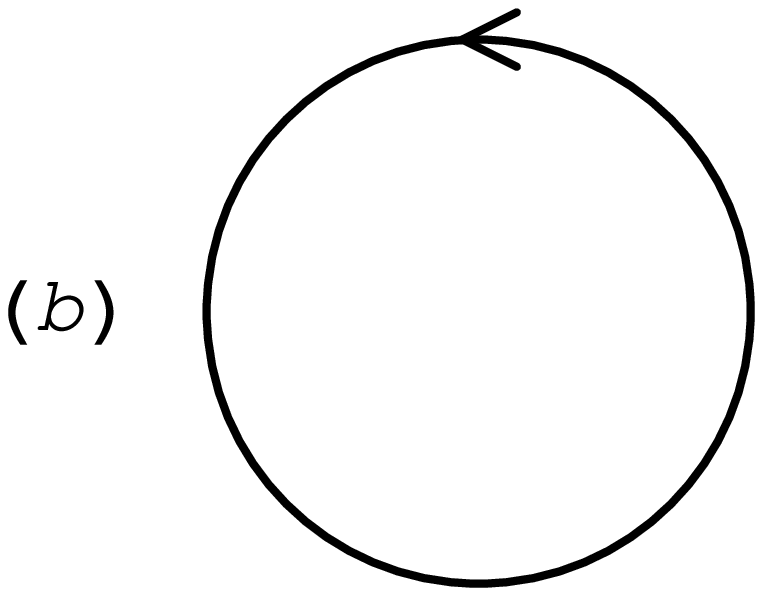}} \\ 
\resizebox{30mm}{!}{\includegraphics{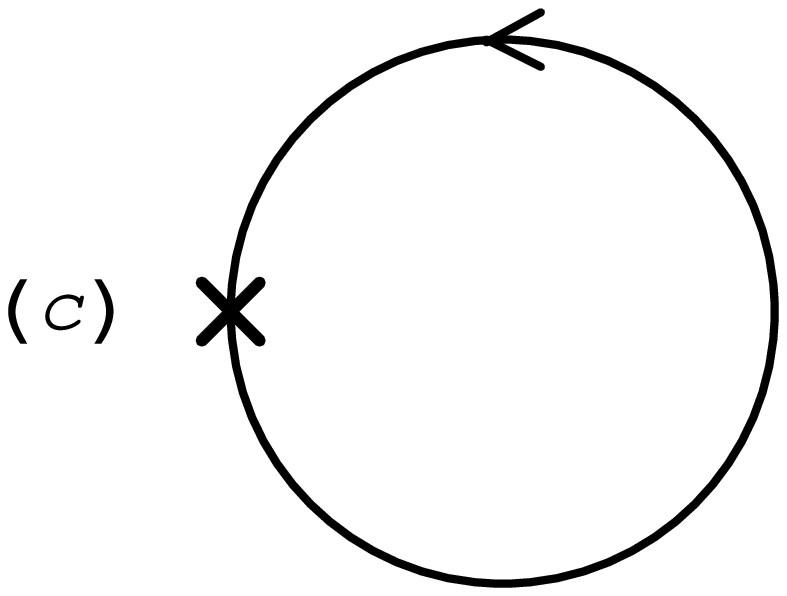}} & \quad %
\resizebox{30mm}{!}{\includegraphics{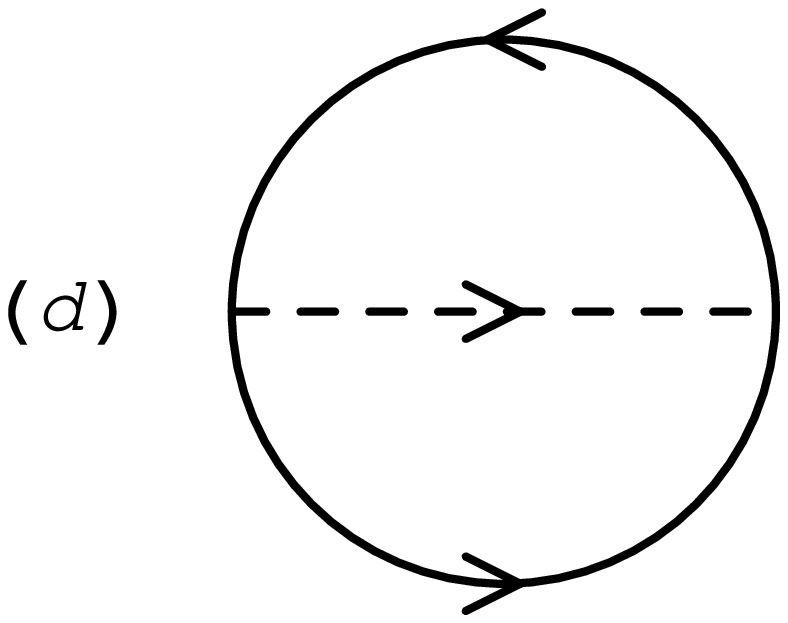}} \\ 
& 
\end{tabular}%
\end{center}
\caption{Feynman diagrams for the $\protect\phi _{0}$ effective potential up
to $\mathcal{O}(\protect\epsilon )$. The solid triangle is the $\protect\phi %
_{0}^{2}/c_{0}$ potential. The solid lines are fermion propagators and
dashed lines are boson propagators. The cross denotes one insertion of $%
\protect\mu _{B}$.}
\end{figure}


\subsection{Near the Unitarity Limit}

Now we study $\mu /\varepsilon _{F}$ near the unitary limit by small $B$
expansion of $V_{\mathrm{eff}}$ to $\mathcal{O}(B)$. The $\mathcal{O}%
(\epsilon ^{-1})$ contributions cancel as 
\begin{equation}
-V_{c_{0}}(\phi _{0})=V_{\mathrm{1L}}(\phi _{0})=\frac{m^{2}\phi _{0}^{2}}{%
8\pi ^{2}\epsilon }B+\mathcal{O}(\epsilon ^{0}).
\end{equation}

The resulting $\overline{\phi }_{0}$ that minimizes $V_{\mathrm{eff}}$ is 
\begin{eqnarray}
\overline{\phi }_{0} &=&\frac{\mu _{B}}{\epsilon }+\frac{B}{2}\left( 2Q(%
\overline{\phi }_{0})+1\right)  \notag \\
&+&\left[ \frac{\mu _{B}}{\epsilon }(3C_{1}+\ln 2-1)+B\left\{ C_{2}+\left(
3C_{1}+\ln 2-\frac{1}{2}\right) \left( Q(\overline{\phi }_{0})+\frac{1}{2}%
\right) -\frac{1}{4}+\frac{\pi ^{2}}{48}\right\} \right] \epsilon +\mathcal{O%
}(\epsilon ^{2}),  \label{U1}
\end{eqnarray}%
where $C_{1}=\tilde{C}\left( 0\right) =0.1442$ \cite{NS1} and $C_{2}=\tilde{C%
}^{\prime }\left( 0\right) =-0.223$ arise from $V_{\mathrm{2L}}(\phi _{0})$
with 
\begin{equation}
V_{\mathrm{2L}}(\phi _{0})=-\epsilon \left( C_{1}\phi _{0}+C_{2}\frac{B}{2}%
\right) \left( \frac{\phi _{0}m}{2\pi }\right) ^{\frac{4-\epsilon }{2}}+%
\mathcal{O}(\epsilon ^{2}).
\end{equation}%
Since $\overline{\phi }_{0}\sim \mathcal{O}(\epsilon ^{0})$, Eq.~(\ref{U1})
shows that $\mu _{B}$ should be $\mathcal{O}(\epsilon )$, which agrees with
the intuitive argument given above.

Next, we obtain $\varepsilon _{F}$ as a function of $\mu _{B}$: 
\begin{equation}
\frac{\mu _{B}}{\epsilon \varepsilon _{F}}=-\frac{B}{2\varepsilon _{F}}%
\left( 2Q(\overline{\phi }_{0})+1\right) +\epsilon ^{1/2}-\frac{B}{%
\varepsilon _{F}}\left( C_{2}+\frac{Q(\overline{\phi }_{0})}{2}+\frac{\pi
^{2}}{48}-\frac{1}{4}\right) \epsilon +\mathcal{O}(\epsilon ^{5/2}).
\label{U2}
\end{equation}%
By taking the LO of $\overline{\phi }_{0}$ in Eq.~(\ref{U1}) and
substituting $\mu _{B}$ by the LO expression in Eq.~(\ref{U2}), one obtains $%
\overline{\phi }_{0}=\varepsilon _{F}\epsilon ^{1/2}+\mathcal{O}(\epsilon )$
which implies $\varepsilon _{F}\sim \mathcal{O}(\epsilon ^{-1/2})$. $%
\overline{\phi }_{0}$ can further be solved to higher orders in $\epsilon $
by interaction. Finally, we obtain 
%
\begin{eqnarray}
\frac{\mu }{\varepsilon _{F}} &=&\frac{\epsilon ^{3/2}}{2}\left[ 1+\frac{%
\epsilon \ln \epsilon }{8}+\frac{5(1-\ln {2})-12C_{1}}{4}\epsilon \right] -%
\frac{1}{2}\left[ sign[a]\left( \frac{B}{\varepsilon _{F}\epsilon ^{1/2}}%
\right) ^{-\epsilon /2}+\frac{\epsilon }{2}\right] \frac{B}{\varepsilon _{F}}%
+\mathcal{O}(\epsilon ^{7/2},B\epsilon ^{5/2})  \notag \\
&\rightarrow &0.475-0.707\eta -0.5\eta ^{2},  \label{NUL1}
\end{eqnarray}%
where the arrow in the second line denotes extrapolation to three dimension $%
\epsilon \rightarrow 1$, and we have used $B/\varepsilon
_{F}=2(ak_{F})^{-2}=2\eta ^{2}$. Note that the result of Ref. \cite{NS2} is
reproduced if we truncate the expansion of $\left( B/\varepsilon
_{F}\epsilon ^{1/2}\right) ^{-\epsilon /2}$ as $1-\epsilon /2\log
[B/\varepsilon _{F}]+\epsilon /4\log [\epsilon ]$, which does not give a
linear term in $\eta $. In order to compare with the Quantum Monte Carlo
(QMC) results \cite{QMC1,QMC2}, $\mu /\varepsilon _{F}$ can be converted to
energy per particle relative to the value of a non-interacting fermi gas 
\begin{equation}
\xi =\frac{{E/A}}{{E_{0}/A}}\ 
\end{equation}%
with $E_{0}/A=3/5\varepsilon _{F}$ in 3-d. The relation is, $\mu
/\varepsilon _{F}=\xi -\frac{\eta }{5}\frac{\partial \xi }{\partial \eta }$.
Our result gives $\xi \simeq 0.475-0.884\eta -0.833\eta ^{2}$. This is
compared with two sets of QMC results denoted as QMC1\cite{QMC1} and QMC2 
\cite{QMC2} in Fig.~2. 

\begin{figure}[tbp]
\begin{center}
\begin{tabular}{c}
\resizebox{100mm}{!}{\includegraphics{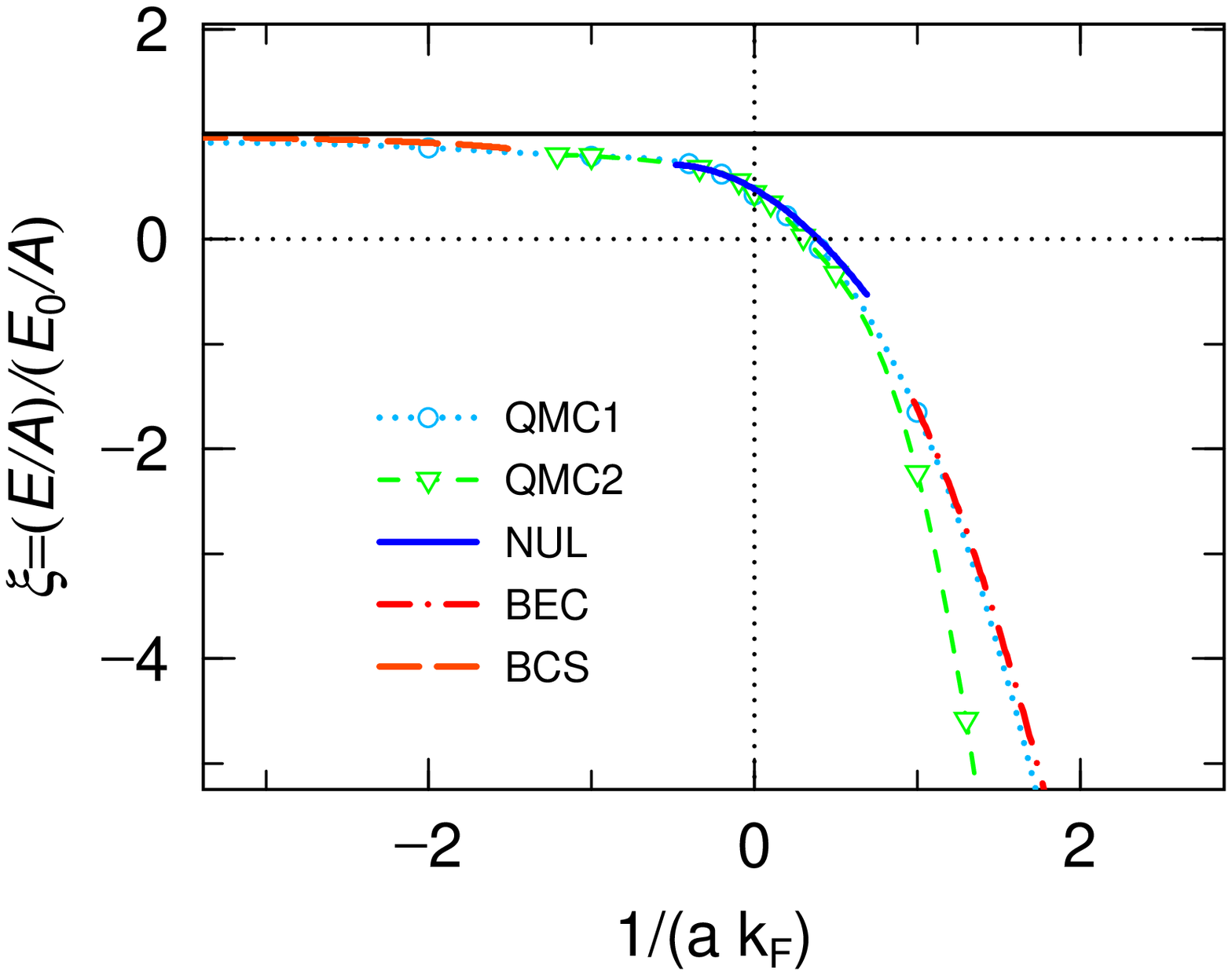}} \\ 
\resizebox{100mm}{!}{\includegraphics{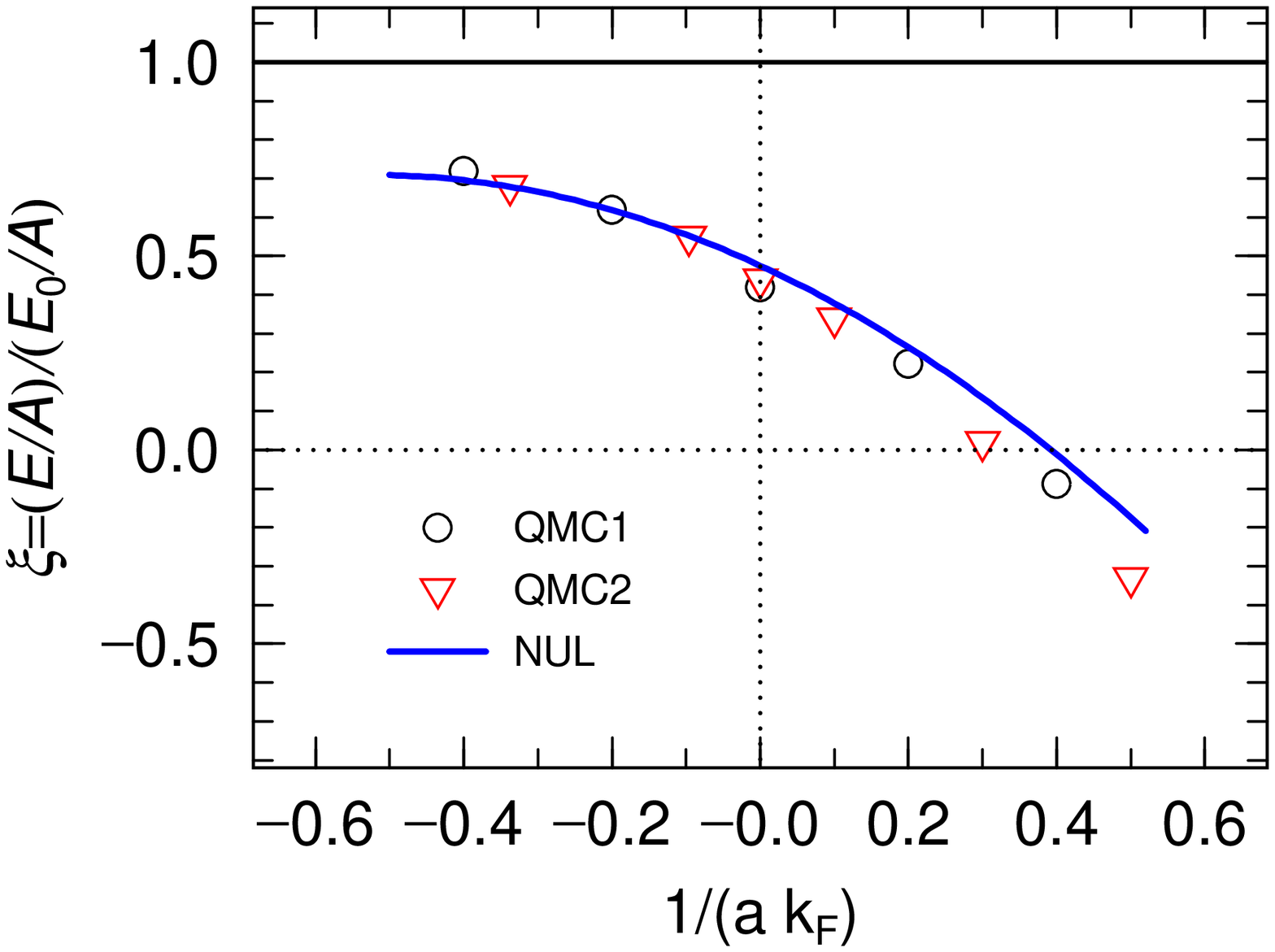}} \\ 
\end{tabular}%
\par
\end{center}
\caption{(Color online) $\protect\xi $ (energy per particle relative to the
value of a non-interacting fermi gas) shown as functions of $(ak_{F})^{-1}$.
QMC1 \protect\cite{QMC1} and QMC2 \protect\cite{QMC2} are Quantum Monte
Carlo calculations with zero and finite range, respectively. NUL is our near
unitarity limit result of Eq. (\protect\ref{NUL1}) in the $\protect\epsilon $
expansion. BEC is our result in Eq. (\protect\ref{BEC1}). BCS is obtained
from numerical solving Eq.~(\protect\ref{EA1}). The lower panel is the
blow-up of the near unitarity regime. }
\end{figure}


\subsection{The BEC Limit}

In BEC limit, $B\gg \varepsilon _{F}$ or $\eta \gg 1$, we perform $1/B$
expansion to $\mathcal{O}(B^{-2})$. Note that $V_{\mathrm{2L}}(\phi
_{0})\sim \mathcal{O}(B^{-2})$ and is neglected. The effective potential has
a simple expression 
\begin{equation}
V_{\mathrm{eff}}\left( \phi _{0}\right) =\frac{2^{\epsilon -8}m^{2}\pi
^{\epsilon /2-1}\phi _{0}^{2}\left( \epsilon \phi _{0}^{2}-16B\mu
_{B}\right) }{B(Bm)^{\epsilon /2}\sin \left( \frac{\pi \epsilon }{2}\right)
\Gamma \left( 1-\frac{\epsilon }{2}\right) }+\mathcal{O}(B^{-2}),
\end{equation}%
which is minimized when $\phi _{0}=\overline{\phi }_{0}=2\sqrt{2B\mu
_{B}/\epsilon }$. Following the same procedure, we obtain 
\begin{eqnarray}
\frac{\mu }{\varepsilon _{F}} &=&-\frac{B}{2\varepsilon _{F}}+\frac{\epsilon
^{2}}{16}\left( 1+\frac{3-\ln {4}}{4}\epsilon \right) \left( \frac{B}{%
\varepsilon _{F}}\right) ^{\epsilon /2-1}  \notag \\
&\rightarrow &0.062\eta ^{-1}-\eta ^{2}.  \label{BEC1}
\end{eqnarray}%
This result is corresponding to $\xi =0.052\eta ^{-1}-\frac{5}{3}\eta ^{2}$.
Note that the $\eta ^{2}$ term just comes from the free molecule
contributions, while the $\eta ^{-1}$ correction is from the $\phi _{0}^{4}$
term in $V_{\mathrm{eff}}$, corresponding to the mean boson-boson
interaction. The numerical value agrees well with the QMC1 result, which
yields $\xi =0.055\eta ^{-1}-\frac{5}{3}\eta ^{2}$ \cite{QMC2}. The QMC2
calculation used a finite range $\cosh $ shape potential, thus the deviation
from the QMC1 curve indicate the finite range effects.


\subsection{Power Counting in the BCS Side}

Next we will turn to the BCS limit, $-\eta \gg 1$, where the situation
changes drastically. As $-a\rightarrow 0$, the gap vanishes and the ground
state should be a non-interacting Fermion gas with $\Delta \equiv \overline{%
\phi }_{0}/\mu \ll 1$. Thus we will count $\mu $ to be $\mathcal{O}(\epsilon
^{0})$. And since Eq.~(\ref{eq1}) is no longer valid, 
we will write the Lagrangian as $L=L_{0}+L_{1}+L_{2}$, where 
\begin{eqnarray}
L_{0} &=&\Psi ^{\dagger }\left[ i\partial _{t}+\left( \frac{\nabla ^{2}}{2m}%
+\mu \right) \sigma _{3}+\phi _{0}\sigma _{+}+\phi _{0}\sigma _{-}\right]
\Psi +\varphi ^{\ast }\left( i\partial _{t}+\frac{\nabla ^{2}}{4m}+2\mu -%
\frac{g^{2}}{c_{0}}\right) \varphi -\frac{\phi _{0}^{2}}{c_{0}},  \notag \\
L_{1} &=&g\Psi ^{\dagger }\sigma _{+}\Psi \varphi +g\Psi ^{\dagger }\sigma
_{-}\Psi \varphi ^{\ast },  \notag \\
L_{2} &=&-\varphi ^{\ast }\left( i\partial _{t}+\frac{\nabla ^{2}}{4m}+2\mu
\right) \varphi -g\frac{\phi _{0}}{c_{0}}(\varphi ^{\ast }+\varphi ).
\end{eqnarray}%
In $L_{0}$ the boson has an \textquotedblleft effective
mass\textquotedblright\ $g^{2}/c_{0}=B+\mathcal{O}(\epsilon )$ which will
suppress the boson propagation in the BCS limit where $B=1/ma^{2}$ is big.
Again, the boson self-energy renormalization vanishes at 4-d.$\mathcal{\ }$%
The effective potential receives leading contributions from Fig.~1(a), (b),
and (d) and yields 
\begin{eqnarray}
V_{\mathrm{eff}}(\phi _{0}) &=&V_{c_{0}}(\phi _{0})-\int \frac{\mathrm{d}%
p^{5-\epsilon }}{i(2\pi )^{5-\epsilon }}\ln \det \left[ S_{F}^{-1}(p,-2\mu )%
\right]   \notag \\
&&+g^{2}\int \frac{\mathrm{d}p^{5-\epsilon }\,\mathrm{d}q^{5-\epsilon }}{%
i(2\pi )^{5-\epsilon }i(2\pi )^{5-\epsilon }}\mathrm{Tr}\left[ S_{F}(p,-2\mu
)\sigma _{+}S_{F}(q,-2\mu )\sigma _{-}\right] \widetilde{D}(p-q),
\label{VBCS1}
\end{eqnarray}%
where $S_{F}(p,-2\mu )$ is the one defined in Eq.(\ref{FP1}) with the
substitution $B\rightarrow -2\mu $. The boson propagator $\widetilde{D}%
^{-1}(p)=-p_{0}(1+i\delta )+\varepsilon _{p}/2+B-2\mu $ has been replaced by 
$\widetilde{D}^{-1}(p)=B$ which is a good approximation when $B\gg 2\mu $ or
equivalently $-\eta \gg 1$. Since we expect $\Delta \ll 1$, by expanding to $%
\mathcal{O}(\Delta ^{2})$ we obtain 
\begin{eqnarray}
\frac{V_{\mathrm{eff}}}{m^{2-\epsilon /2}\mu ^{3-\epsilon /2}} &=&-\frac{1}{%
12\pi ^{2}}\left[ 1+\frac{\epsilon }{2}\left( 3\tilde{B}^{\epsilon /2-1}+\ln
(2\pi )-\gamma +\frac{11}{6}\right) \right] +\frac{\Delta ^{2}}{8\pi ^{2}}%
\left[ \frac{\tilde{B}^{1-\epsilon /2}-2}{\epsilon }\right.   \notag \\
&&+\left. \frac{1-\gamma +\ln \left( 4\pi \right) }{2}\tilde{B}^{1-\epsilon
/2}+2\tilde{B}^{\epsilon /2-1}+\gamma +\ln \frac{\Delta ^{2}}{8\pi }\right] ,
\label{VBCS2}
\end{eqnarray}%
where $\tilde{B}\equiv B/\mu $.

At this order, the gap equation, $\partial V_{\mathrm{eff}}/\partial \Delta
=0$, has a simple solution 
\begin{equation}
\epsilon \ln \Delta =1-\frac{\tilde{B}^{1-\epsilon /2}}{2}-\epsilon \left( 
\frac{\ln (4\pi )-\gamma +1}{4}\tilde{B}+\tilde{B}^{-1}+\frac{1+\gamma -\ln
(8\pi )}{2}\right) +\mathcal{O}(\epsilon ^{2},\Delta ^{2}).  \label{BCSSol}
\end{equation}%
Thus, $\ln \Delta \sim \mathcal{O}(\epsilon ^{-1})$.\ We will formally count 
$\Delta \sim \mathcal{O}(\epsilon ^{1/2})$ in Eq.(\ref{VBCS2}) to indicate
the smallness of the gap.

The $n_{f}$ and $\varepsilon _{F}$ can be computed using Eqs.~(\ref{a}) and (%
\ref{b}): 
\begin{eqnarray}
\frac{\mu }{\varepsilon _{F}} &=&1-\epsilon \tilde{B}^{\epsilon /2-1}-\frac{%
3\Delta ^{2}}{2\epsilon }\left\{ 1-\frac{\tilde{B}^{1-\epsilon /2}}{3}\right.
\notag \\
&&+\left. \epsilon \left[ \tilde{B}\left( \frac{6\gamma -6\ln (8\pi )-11}{72}%
\right) -\frac{7}{3}\tilde{B}^{-1}+\frac{1-2\gamma +10\ln 2+2\ln \pi }{8}%
-\ln \Delta \right] \right\} .  \label{MUEP1}
\end{eqnarray}%
Since $\tilde{B}=\frac{2}{\left( ak_{F}\right) ^{2}}\frac{\varepsilon _{F}}{%
\mu }$ and $\mu /\varepsilon _{F}=1+\mathcal{O}(\epsilon ^{0})$, Eq.~(\ref%
{BCSSol}) shows that 
\begin{equation}
\Delta \simeq \exp \left[ \frac{1-|ak_{F}|^{-2+\epsilon }}{\epsilon }\right]
.
\end{equation}%
This implies that, in 4-d, the gap vanishes and $\mu /\varepsilon _{F}=1$
when $-ak_{F}\ll 1$. In 3-d and with $-ak_{F}\ll 1$, the gap is
exponentially suppressed. One might argue that $\Delta $ should be smaller
than $\mathcal{O}(\epsilon ^{1/2})$ as counted in Eq.(\ref{VBCS2}). This is
certainly true when $\epsilon \rightarrow 0$. However, since we will take $%
\epsilon \rightarrow 1$ at the end, this counting suits our purpose
especially when working between the BCS and unitarity limits.

To compare with the low energy theorem $\xi =1+\frac{10}{9\pi }\eta ^{-1}+%
\mathcal{O}(\eta ^{-2})$ obtained in \cite{KH1,TD1}, we perform a $1/\eta $
expansion to Eq.(\ref{MUEP1}). At NLO ($\mathcal{O}(\epsilon )$) with all
the exponentially suppressed $\mathcal{O}(\Delta ^{2})$ terms neglected, we
see no $\eta ^{-1}$ correction. However, we expect that at NNLO, the $%
\epsilon \tilde{B}^{\epsilon /2-1}$ will be fully recovered and gives 
\begin{equation}
\frac{\mu }{\varepsilon _{F}}\simeq 1+\frac{ak_{F}}{\sqrt{2}},
\end{equation}%
as $\epsilon \rightarrow 1$. This implies $\xi =1+0.589\eta ^{-1}+\mathcal{O}%
(\eta ^{-2},\epsilon ^{2})$. This agrees reasonably well with the low energy
theorem.

As $-\eta $ becomes smaller but not much smaller than unity, the $\mathcal{O}%
(\Delta ^{2})$ effect in $\mu /\varepsilon _{F}$ becomes more important. 
From the thermodynamic relation $E/A=V_{\mathrm{eff}}/n_{f}+\mu $, we obtain

\begin{eqnarray}
\xi &=&1-\epsilon \frac{3}{4}\tilde{B}^{-1}  \notag \\
&&+\frac{1}{8}\left[ 27\tilde{B}^{-1}+\left[ 3\gamma -3\ln (2\pi )-4\right] 
\tilde{B}-13-6\gamma +6\ln (2\pi )\right] \Delta ^{2}.  \label{EA1}
\end{eqnarray}%
Eqs.~(\ref{BCSSol}) and (\ref{EA1}) give $\xi $ in terms of $\tilde{B}$,
while Eq.(\ref{MUEP1}) gives $\tilde{B}$ in terms of $ak_{F}$. The numerical
result of $\xi $ as a function of $ak_{F}$ is shown as the BCS curve in
Fig.~2.

\section{Conclusions}

We have extended the $\epsilon $ expansion (expansion around four spacial
dimensions) developed by Nishida and Son for a cold fermi gas with infinite
scattering length to finite scattering length to study the BEC-BCS
crossover. A resummation of higher order logarithms and a suitable extension
of fermion coupling in d-dimensions\ (see the discussion around Eqs. (\ref%
{C0x}-\ref{resum})) have been developed in order to apply the theory in the
BCS regime. The ratio between the chemical potential and the Fermi energy, $%
\mu /\varepsilon _{F}$, has been computed to next-to-leading order in the $%
\epsilon $ expansion as a function of $\eta =1/(ak_{F})$. Near the unitarity
limit $\left\vert \eta \right\vert \rightarrow 0$, we found $\mu
/\varepsilon _{F}=0.475-0.707\eta -0.5\eta ^{2}$. Near the BEC limit $\eta
\rightarrow \infty $, $\mu /\varepsilon _{F}=0.062/\eta -\eta ^{2}$, and
near the BCS limit $\eta \rightarrow -\infty $, $\mu /\varepsilon
_{F}=1+0.707/\eta $. As shown in Fig.~2, overall good agreement with Quantum
Monte Carlo results has been found.

\section{Appendix}

The function $\tilde{C}\left( z\right) $ that appears in Eq. (\ref{Cz}) is
defined as 
\begin{eqnarray*}
\tilde{C}\left( z\right) &=&\int_{z/2}^{\infty }dx\int_{z/2}^{\infty }dy%
\frac{\left( \sqrt{x^{2}+1}-x\right) \left( \sqrt{y^{2}+1}-y\right) }{2\sqrt{%
x^{2}+1}\sqrt{y^{2}+1}} \\
&&\times \left( -z+x+y+2\sqrt{x^{2}+1}+2\sqrt{y^{2}+1}-\sqrt{\left[
-z+x+y+2\left( \sqrt{x^{2}+1}+\sqrt{y^{2}+1}\right) \right] ^{2}-(z-2x)(z-2y)%
}\right) .
\end{eqnarray*}%
The numerical value of $\tilde{C}\left( z\right) $ is plotted in Fig.~3. 
\begin{figure}[tbp]
\begin{center}
\begin{tabular}{c}
\resizebox{75mm}{!}{\includegraphics{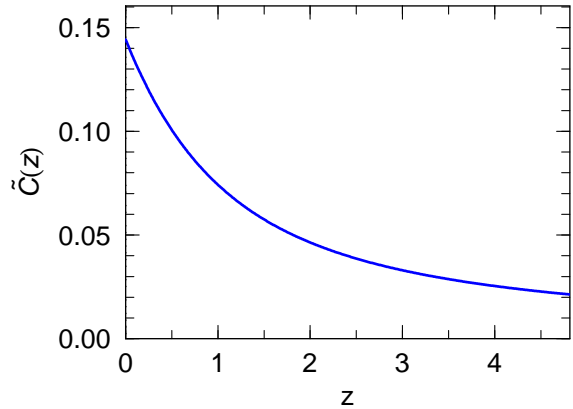}} \\ 
\end{tabular}%
\end{center}
\caption{The $\tilde{C}(z)$ function that appears in Eq. (\protect\ref{Cz})}
\label{FIG3}
\end{figure}

In Eq.(\ref{VBCS1}), the one fermion loop contribution to the effective
potential near the BCS limit has the analytic expression:

\begin{multline*}
-\int \frac{\mathrm{d}p^{5-\epsilon }}{i(2\pi )^{5-\epsilon }}\ln \det \left[
S_{F}^{-1}(p,-2\mu )\right] =\frac{2^{\frac{\epsilon }{2}-4}\pi ^{\frac{%
\epsilon -5}{2}}\Delta ^{2-\frac{\epsilon }{2}}}{\Gamma \left( 2-\frac{%
\epsilon }{2}\right) } \\
\times \left[ \Delta \Gamma \left( 1-\frac{\epsilon }{4}\right) \Gamma
\left( \frac{\epsilon -6}{4}\right) \,_{2}F_{1}\left( \frac{\epsilon -6}{4},%
\frac{\epsilon -2}{4};\frac{1}{2};-\frac{1}{\Delta ^{2}}\right) +2\Gamma
\left( \frac{3}{2}-\frac{\epsilon }{4}\right) \Gamma \left( \frac{\epsilon }{%
4}-1\right) \,_{2}F_{1}\left( \frac{\epsilon }{4},\frac{\epsilon }{4}-1;%
\frac{3}{2};-\frac{1}{\Delta ^{2}}\right) \right] 
\end{multline*}



\section{Acknowledgements}

We would like to thank Chung-Wen Kao for helpful discussion. This work was
supported by the NSC and NCTS of Taiwan, ROC.


\end{document}